\documentclass[conference]{IEEEtran}
\IEEEoverridecommandlockouts

\usepackage[T1]{fontenc}

\usepackage{cite}
\usepackage{amsmath,amssymb,amsfonts}
\usepackage{algorithmic}
\usepackage{graphicx}
\usepackage{textcomp}
\usepackage[svgnames,table]{xcolor}

\usepackage[svgnames,table]{xcolor}
\definecolor{darkblue}{rgb}{0, 0, 0.7}

\usepackage{verbatim}

\usepackage{amsfonts}
\usepackage{amssymb}
\usepackage{amsmath}
\usepackage{times}

\usepackage{todonotes}

\usepackage{caption}
\usepackage{subcaption}
\usepackage{pifont}
\usepackage{booktabs}

\usepackage{wrapfig}

\usepackage{paralist}

\usepackage{listings}
\usepackage[colorlinks=true,
pagebackref=false,
urlcolor=blue!50!blue,
citecolor=blue!50!blue,
linkcolor=darkblue]{hyperref}

\usepackage[capitalise,nameinlink]{cleveref}
\crefname{lstlisting}{listing}{listings}
\Crefname{lstlisting}{Listing}{Listings}
\crefname{line}{Line}{Lines}
\Crefname{line}{L}{L}
\crefname{section}{§}{§§}
\Crefname{section}{§}{§§}

\usepackage{xspace} 

\newcommand{\eg}{e.g.\xspace}

\newcommand{\ie}{i.e.\xspace}
\newcommand{\st}{s.t.\xspace}

\definecolor{badred}{RGB}{240,0, 0}
\definecolor{goodgreen}{RGB}{0, 216, 0}

\newcommand{\graphlevel}{\emph{Road Network level}\xspace}
\newcommand{\geographiclevel}{\emph{Geometric level}\xspace}
\newcommand{\roadlevel}{\emph{Road level}\xspace}
\newcommand{\laneletlevel}{\emph{Navigation level}\xspace}
\newcommand{\fulldetail}{\emph{Full Detail level}\xspace}

\usepackage[acronyms,shortcuts,section=section,style=super,nopostdot,nogroupskip,nomain,nonumberlist]{glossaries}
\glsdisablehyper

\newacronym{ads}    {ADS}   {automated driving system}
\newacronym{av}     {AV}    {autonomous vehicle}

\newacronym{gis}    {GIS}   {geographic information system}

\newacronym{cas}    {CAS}   {collision avoidance system}
\newacronym{cps}    {CPS}   {cyber-physical system}
\newacronym{ea}     {EA}    {evolutionary algorithm}
\newacronym{ed}     {ED}    {Euclidean distance}
\newacronym{ga}     {GA}    {genetic algorithm}
\newacronym{gd}     {GD}    {generational distance}
\newacronym{hv}     {HV}    {hypervolume}
\newacronym{igd}    {IGD}   {inverse generational distance}
\newacronym{iot}    {IoT}   {internet of things}
\newacronym{knn}    {kNN}   {k-nearest neighbours}
\newacronym{mlsm}   {MLSM}  {multi-level scenario model}
\newacronym{moo}    {MOO}   {multi-objective optimisation}
\newacronym{mop}    {MOP}   {multi-objective problem}
\newacronym{mosp}    {MOSP}   {multi-objective search problem}
\newacronym{osm}    {OSM}   {Open Street Maps}
\newacronym{qi}     {QI}    {quality indicator}
\newacronym{sbt}    {SBT}   {search-based testing}
\newacronym{sdb}    {S-DB}  {scenario database}
\newacronym{sdl}    {SDL}   {scenario description language}

\usepackage{tikz}

\usepackage[clock]{ifsym} 

\usetikzlibrary{positioning}
\usetikzlibrary{calc} 
\usetikzlibrary{arrows}
\usetikzlibrary{arrows.meta}
\usetikzlibrary{shapes}
\usetikzlibrary{arrows.meta}
\usetikzlibrary{patterns}

\definecolor{ao}{rgb}{0.0, 0.5, 0.0}
\pgfmathsetmacro{\markingwidth}{1pt}

\tikzset{
    roadmarking/.style={draw=gray!80,line width=\markingwidth},
    centreline/.style={roadmarking,dashed,dash pattern=on 7pt off 4pt},
    stopline/.style={roadmarking,line width=4pt},
    basevehicle/.style={rectangle,minimum height=0.5cm,minimum width=0.25cm,inner sep=0pt,},
    vehicle/.style={basevehicle,execute at begin node={\includegraphics[trim=0.5cm 0 0.5cm 0,clip,width=.25cm]{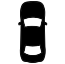}}},
    parkedvehicle/.style={basevehicle,execute at begin node={\includegraphics[trim=0.5cm 0 0.5cm 0,clip,width=.25cm]{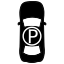}}},
    ego/.style={basevehicle,execute at begin node={\includegraphics[trim=0.5cm 0 0.5cm 0,clip,width=.25cm]{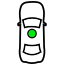}}},
    pedestrian/.style={minimum width=0.2cm, minimum height=0.2cm,inner sep=0cm,execute at begin node={\includegraphics[width=.3cm]{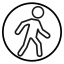}}},
    egopath/.style={draw=blue,-latex,thick},
    vehiclepath/.style={draw=red,-latex,thick},
    pedestrianpath/.style={draw=ao,-latex,thick},
    crosswalk/.style={rectangle,fill=gray!80,minimum height=0.06cm,minimum width=0.35cm,inner sep=0},
    stop/.style={minimum height=0.1cm,execute at begin node={\includegraphics[trim=0.5cm 0 0.5cm 0,clip,width=.25cm]{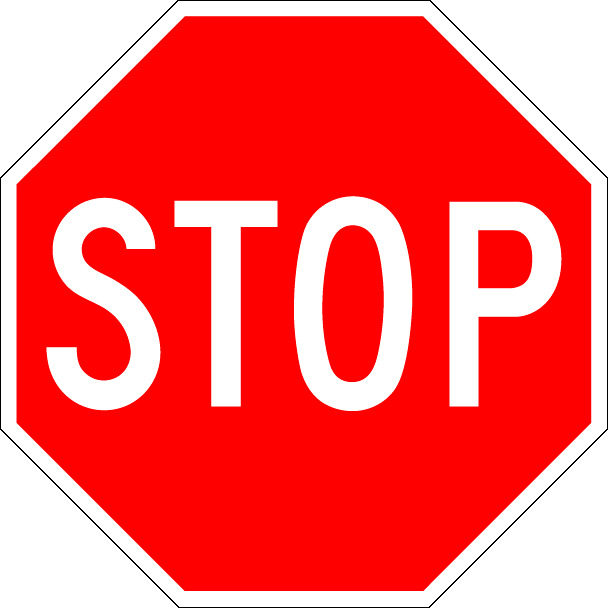}}},
    bump/.style={minimum height=0.1cm,execute at begin node={\includegraphics[trim=0.5cm 0 0.5cm 0,clip,width=.25cm]{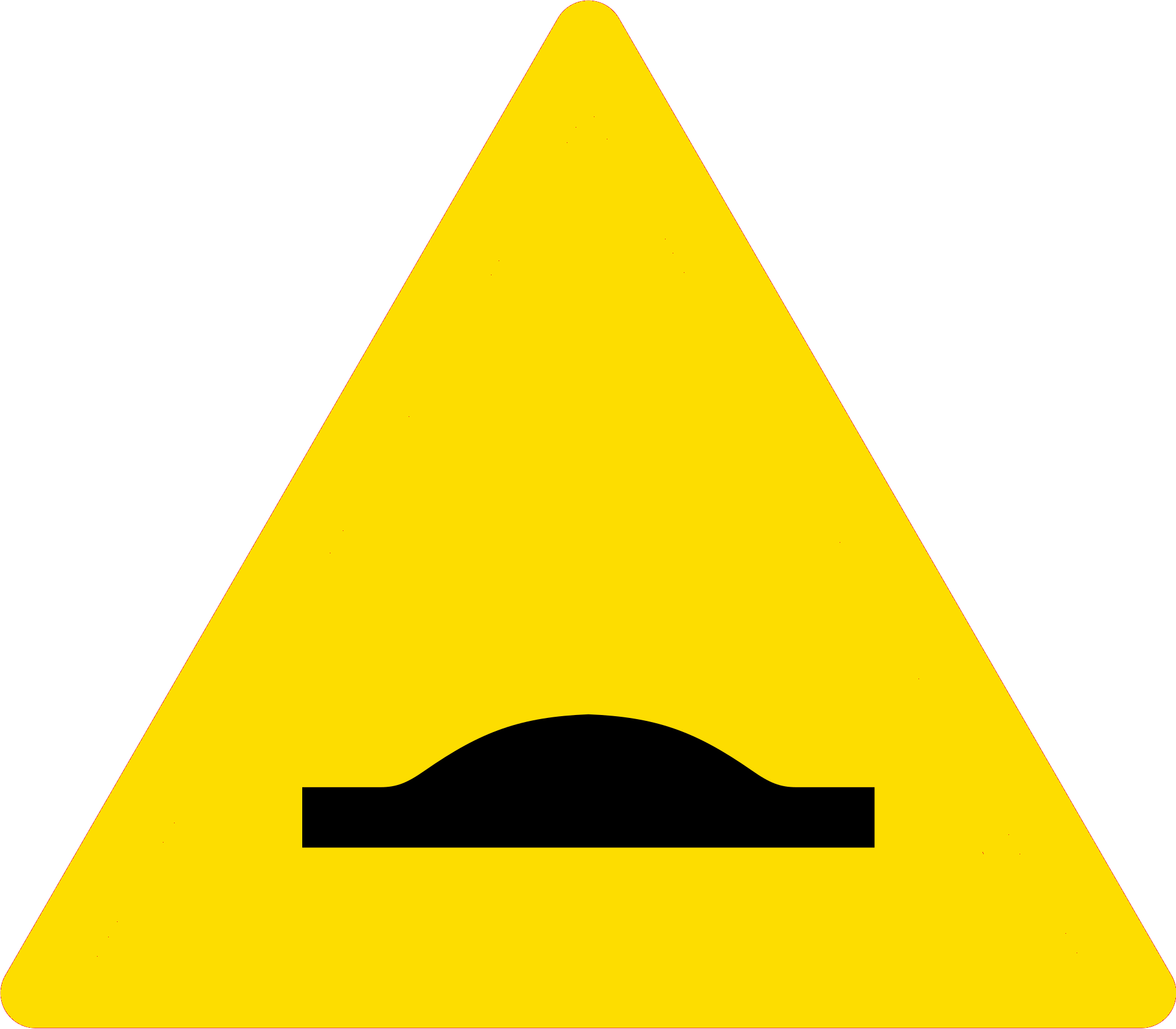}}},
    trafficlight/.style={minimum height=0.1cm,execute at begin node={\includegraphics[trim=0.5cm 0 0.5cm 0,clip,height=.45cm]{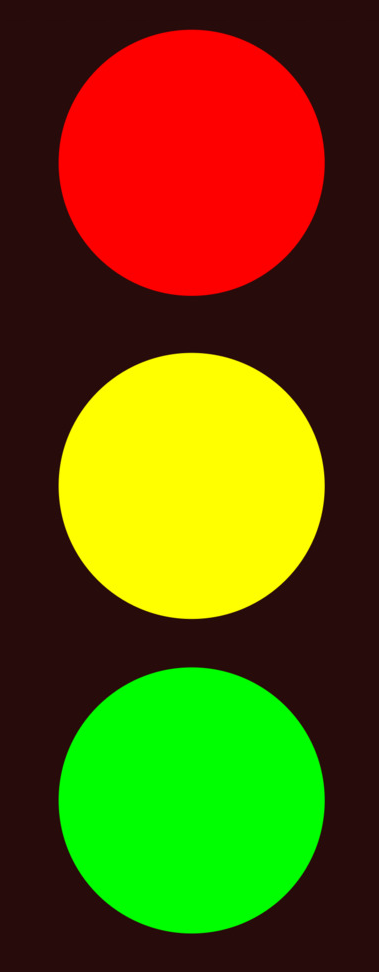}}},
}

\usepackage{background}
\SetBgContents{Preprint}
\SetBgColor{gray}
\SetBgOpacity{0.2}
\SetBgScale{14}
\SetBgAngle{45}
\usepackage{tcolorbox}
\usepackage{fancyvrb}

\begin{document}

\begin{figure*}[ht]  
{
\Large \textbf{PREPRINT}.\\
\\
This is an author preprint version of the paper.\\
\\
See the MPM4CPS'21 proceedings for the final version.
\\
}

\begin{tcolorbox}[title=Cite as,colback=white,left=0pt,right=0pt]
Klikovits S., Arcaini P. (2021) On the Need for Multi-Level ADS Scenarios. In: Proc. 3rd International Workshop on Multi-Paradigm Modelling for Cyber-Physical Systems (MPM4CPS'21). ACM
\end{tcolorbox}

\begin{tcolorbox}[title=Bibtex,colback=white,left=0pt,right=0pt]
\begin{verbatim}
@InProceedings{KlikovitsA2021MPM4CPS,
  title="On the Need for Multi-Level ADS Scenarios",
  year="2021",
  author="Klikovits, Stefan and Arcaini, Paolo",
  booktitle="3rd International Workshop on Multi-Paradigm Modelling for
Cyber-Physical Systems (MPM4CPS'21)",
  location="Virtual Event, Japan",
  series="MODELS'21"
  publisher="Association for Computing Machinery",
}
\end{verbatim}
\end{tcolorbox}
\end{figure*}
\pagebreak
\setcounter{page}{1}


\title{On the Need for Multi-Level ADS Scenarios\\
\thanks{The authors are supported by ERATO HASUO Metamathematics for Systems Design Project (No. JPMJER1603), JST. Funding reference number: 10.13039/501100009024 ERATO. S. Klikovits is also supported by Grant-in-Aid for Research Activity Start-up 20K23334, JSPS.}
}

\author{\IEEEauthorblockN{Stefan Klikovits}
\IEEEauthorblockA{
\textit{National Institute of Informatics}\\
Tokyo, Japan \\
klikovits@nii.ac.jp}
\and
\IEEEauthorblockN{Paolo Arcaini}
\IEEEauthorblockA{
\textit{National Institute of Informatics}\\
Tokyo, Japan \\
arcaini@nii.ac.jp}
}

\maketitle

\begin{abstract}
Currently, most existing approaches for the design of \ac{ads} scenarios focus on the description at one particular abstraction level---typically the most detailed one.
This practice often removes information at higher levels, such that this data has to be re-synthesized if needed.
As the abstraction granularity should be adapted to the task at hand, however, engineers currently have the choice between re-calculating the needed data or operating on the wrong level of abstraction.
For instance, the search in a scenario database for a driving scenario with a map of a given road-shape should abstract over the lane markings, adjacent vegetation, or weather situation.
Often though, the general road shape has to be synthesized (\eg interpolated) from the precise GPS information of road boundaries.
This paper outlines our vision for multi-level \ac{ads} scenario models that facilitate scenario search, generation, and design.
Our concept is based on the common modelling philosophy to interact with scenarios at the most appropriate abstraction level. 
We identify different abstraction levels of \ac{ads} scenarios and suggest a template abstraction hierarchy. 
Our vision enables seamless traversal to such a most suitable granularity level for any given scenario, search and modelling task.
We envision that this approach to \ac{ads} scenario modelling will have a lasting impact on the way we store, search, design, and generate \ac{ads} scenarios, allowing for a more strategic verification of autonomous vehicles in the long run.
\end{abstract}

\begin{IEEEkeywords}
Automated Driving Systems, Scenario Model, Hierarchical Modeling
\end{IEEEkeywords}

\glsresetall


\section{Introduction}
At a time when companies such as Tesla, Baidu and Waymo continuously release newer generations of self-driving vehicles onto public roads, the need for proper verification and validation of \acp{ads} is undeniable.
As a pure hardware-in-the-loop verification is practically infeasible~\cite{Kalra2016DrivingTS}, the de-facto standard is to develop, test and verify the \ac{ads}'s decision-making software in simulated environments. 
Next to typical benefits such as reduced cost, parallelisation and faster-than-reality simulation speeds, such simulation-based verification is of particular interest to test the \ac{ads}' behaviour in situations that in reality only rarely occur~\cite{sarkar2019behavior}.

Scenarios, typically captured in some form of \ac{sdl} (\eg CommonRoad~\cite{Althoff2017CommonRoadCB}, OpenSCENARIO~\cite{openscenariov2}, GeoScenario~\cite{queiroz2019geoscenario}), provide the necessary means to capture essential simulation details and describe the following five dimensions~\cite{IlievskiMarko2020}:
\begin{inparaenum}
\item road structure and geometry,
\item traffic interactions, 
\item static objects, 
\item weather conditions, and 
\item occlusions.
\end{inparaenum}
The \acp{sdl}' difference lies in the supported granularity for the individual dimensions, \st some enable a more precise description of vehicle behaviour, while others may focus on detailed description of roads or obstacle interaction.
These scenarios, collected in suites, are then used for verification and validation of \ac{ads} software.

Typically, we can distinguish three principal ways of obtaining \ac{ads} scenarios.
\begin{itemize}
\item \emph{Harvesting}, \ie retrieving information from the ``real world'', \eg by extracting sensor and camera data or record \& replay from real vehicle drives; or by using cartography sources such as \ac{osm} to extract road and mapping information.
\item \emph{Generation}, \ie the creation of artificial maps or scenarios according to certain requirements and constraints.
\item \emph{Design}, \ie the creation of scenarios by hand using expert knowledge to create specific situations.
\end{itemize}
Evidently, these techniques can be combined, such that a harvested scenario can be adapted by an engineer by hand. 
Similarly, automated search-based generation techniques can be applied on a manually designed road situation \eg to increase the likelihood of crash situations.

\tikzset{
roadnode/.style={circle,fill=black,inner sep=0pt,minimum size=6pt},
roadedge/.style={draw=gray,very thick}
}

\begin{figure*}[t]
\begin{subfigure}[t]{0.20\textwidth}
\centering
\begin{tikzpicture}
    \node[roadnode,fill=red] (one) {};
    \node[below left=1 and 1 of one] (two) {};
    \node[roadnode,below right=1 and 1 of one] (three) {};
    \node[roadnode,below right=1 and 1 of two] (four) {};
    \draw[-latex,roadedge] (one) -- (two);
    \draw[latex-latex,roadedge] (one) -- (three);
    \draw[latex-latex,roadedge] (one) -- (four);
    \draw[latex-latex,roadedge] (three) -- (four);
    
    \node[rectangle,draw=black,thin,below=1 of three,align=center] (text-a) {\scriptsize highway};
    \draw[dashed,thick] (text-a) -- ($(three)!0.5!(four)$);
    
\end{tikzpicture}
\caption{Road Network Graph of intersections and roads}
\label{fig:graph}
\end{subfigure} %
\hfill %
\begin{subfigure}[t]{0.20\textwidth}
\centering
\begin{tikzpicture}
    \node[roadnode,fill=red] (one) {};
    \node[left=1 of one] (two) {};
    \node[roadnode,below=1 of one] (three) {};
    \node[roadnode,below right=1 and 1 of one] (four) {};
    \draw[-latex,roadedge] (one) to[out=190,in=-10] (two);
    \draw[latex-latex,roadedge] (one) to[out=270,in=70] (three);
    \draw[latex-latex,roadedge] (one) to[out=0,in=80] (four);
    \draw[latex-latex,roadedge] (three) to[out=-20,in=160] (four);
    
    \node[rectangle,draw=black,thin,below right=0.5 and 0 of three,align=center] (text-a) {\scriptsize 4 lanes};
    \draw[dashed,thick] (text-a) -- ($(three)!0.5!(four)$);

    \node[rectangle,draw=black,thin,above right=0.15 and 0.75 of one] (text-b) {\scriptsize 2 lanes};
    \draw[dashed,thick] (text-b) -- (1,-.5);

    \node[rectangle,draw=black,thin,above left=0.25 and 0 of one] (text-c) {\scriptsize 2 lanes};
    \draw[dashed,thick] (text-c) -- ($(one)!0.5!(two)$);

    \node[rectangle,draw=black,thin,below left=0.5 and 0.5 of one] (text-d) {\scriptsize 2 lanes};
    \draw[dashed,thick] (text-d) -- ($(one)!0.5!(three)$);

\end{tikzpicture}
\caption{Geographic road network with annotations}
\label{fig:geographic}
\end{subfigure} %
\hfill %
\begin{subfigure}[t]{0.20\textwidth}
\centering
\begin{tikzpicture}[scale=0.6]
    \draw[rectangle,draw=none,fill=green!70!black] (1,0) rectangle (-1,-2);
    \draw[rectangle,draw=none,fill=green!70!black] (-1,2) rectangle (5,2.85);
    \draw[rectangle,draw=none,fill=green!70!black] (3,0) rectangle (5,-2);
    
    \draw[roadmarking] (-1,0) -- ++(2,0);
    \draw[roadmarking] (3,0) -- ++(2,0);
    \draw[centreline] (-1,1) -- (5,1);
    \draw[roadmarking] (-1,2) -- (5,2);
    
    \draw[roadmarking] (1.0,0) -- ++(0,-2);
    \draw[centreline] (2.0,0) -- ++(0,-2);
    \draw[roadmarking] (3.0,0) -- ++(0,-2);
    
    \node[rectangle,fill=black,minimum width=0.6cm,minimum height=4pt,anchor=north west,inner sep=0pt] at (2,-0.25) {};
    \node[trafficlight] at (3.25,-0.5) {};
    
    \node[rectangle,fill=black,minimum height=0.6cm,minimum width=4pt,anchor=south west,inner sep=0pt] at (0.75,0) {};
    \node[trafficlight,rotate=-90] at (0.5,-0.25) {};
    
    \node[rectangle,fill=black,minimum height=0.6cm,minimum width=4pt,anchor=south west,inner sep=0pt] at (3.25,1) {};
    \node[trafficlight,rotate=90] at (3.5,2.25) {};
\end{tikzpicture}
\caption{Road level includes boundaries, lanes and traffic lights.}
\label{fig:road}
\end{subfigure} %
\hfill %
\begin{subfigure}[t]{0.20\textwidth}
    \centering
    \includegraphics[width=\linewidth]{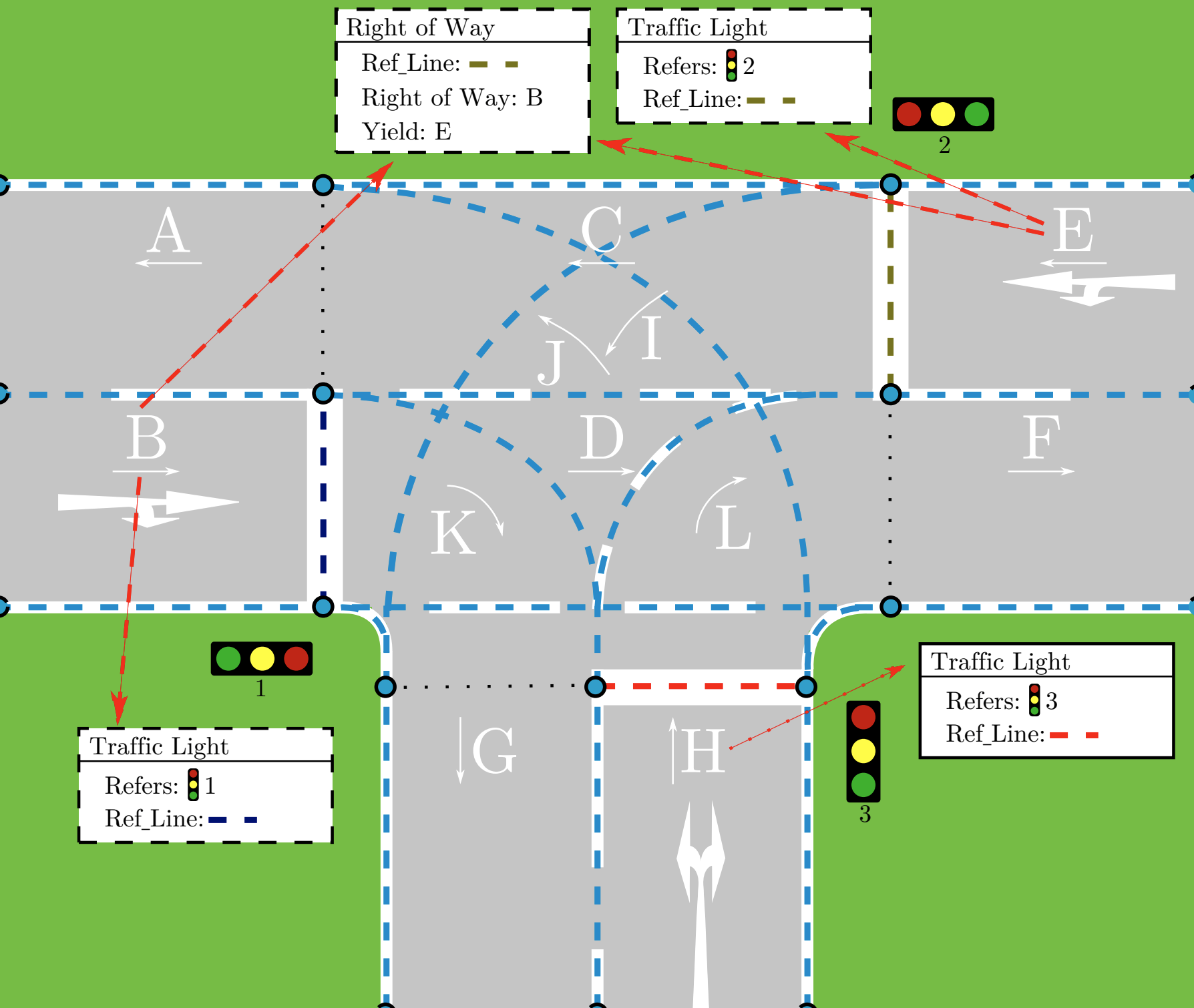}
    \caption{Lanelets example with route information. Image Source: \cite{poggenhans2018lanelet2}}
    \label{fig:lanelets}
\end{subfigure}
\caption{A scenario map on different levels of abstraction. Fig (a) and (b) depict a larger network, Fig (c) and (d) focus on the highlighted crossing \protect\tikz{\protect\node[roadnode,fill=red]{}}.}
\label{fig:fourfigures}
\end{figure*}
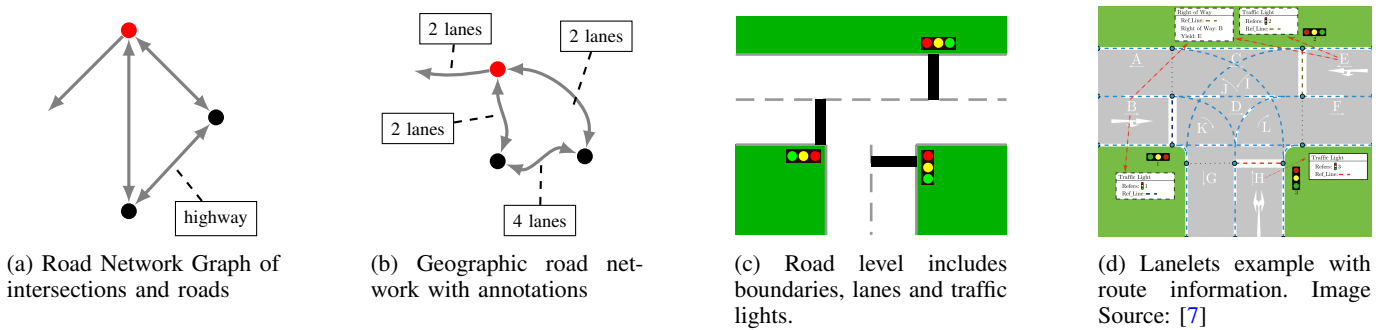

Nonetheless, a scenario's source has a big impact on the abstraction level at which it is captured.
For instance, it is easy to automatically harvest and capture detailed information about road shapes, lane markings and boundaries, and other observations from the sensor and camera recordings, or cartography services.
When designing scenarios by hand, however, the manual editing process is typically very involved, such that it is easier to start at a higher abstraction level, providing only basic map information.
Ideally, semi-automated tools (\eg Matlab's RoadRunner\footnote{\url{https://www.mathworks.com/products/roadrunner.html}}) can be used to generate some of the missing information using \eg existing road geometry standards and conventions, before experts continue to manually edit the scenario, and define more details.

Independent of the source, typical approaches tend to capture scenarios at fine-grained abstraction levels, storing data as detailed as possible.
For instance, CommonRoad and GeoScenario both use lanelets~\cite{poggenhans2018lanelet2} to describe the road geometry.
These fractured, interconnected road lane-pieces split the driving surface and store information such as exact GPS road boundaries, adjoint and successor lanelets, road markings and signs, etc. allowing an \ac{ads} to optimally simulate the situation and realistically plan its trajectory through traffic (see \cref{fig:lanelets}).
As the granularity increases, however, oftentimes high-level information is removed or refined away along the way. 
In our opinion, this is where a problem of this approach lies.

Evidently, the detailed information is ideal for the precise reconstruction and simulation.
Nonetheless, as scenario suites and databases grow, their organisation, querying and search become evermore important.
High-level information about the scenario such as road shape, number of lanes, number and type of non-ego actors, etc. are removed and need to be re-synthesized.
At best, scenario and benchmark databases such as the CommonRoad Benchmarks~\cite{7995802} or the SafetyPool Scenario-Database\footnote{\url{https://www.safetypool.ai/}} allow for some manual keyword-based tagging or textual description, which is clearly insufficient.
We believe that every scenario should include information to dynamically switch to the most appropriate abstraction levels.
This means, querying, analysing and editing should be possible on high abstraction levels (\eg road shape, number of lanes, number of roads leading to crossings), 
and on low granularity levels (exact position of object and road boundaries) likewise.

In light of this vision, we propose the \ac{mlsm}, a framework that organises scenario information on different abstraction levels.
The rest of this paper, introduces the details of this framework (\cref{sec:mlsm}), relating it to typical scenario design use cases. 
Next, we take a step back and describe generic modelling activities and their relation to the framework (\cref{sec:activities}), 
before focusing on one specific use case involving different scenario sources within the framework (\cref{sec:caseStudySBT}).
A specific \ac{sbt} case study shows the advantages of operating on appropriate abstraction levels.
Finally, we share thoughts on a potential implementation (\cref{sec:implementation}) and our next steps leading to realising the \ac{mlsm}.


\section{Multi-Level Scenario Model}
\label{sec:mlsm}

\tikzset{
box/.style={draw,rectangle,rounded corners=0.1em, minimum height=2em, inner sep=0.15em,align=center,minimum width=6em},
source/.style={box,rounded corners=0.25em,draw=none,font=\normalsize},
sourcelevel/.style={box,fill=white},
base/.style={box,thick,fill=gray!10},
refines/.style={-triangle 45,thick,dashed},
myarrow/.style={-latex,thick,dashed},
circled/.style={shape=circle,draw,fill=white,inner sep=2pt,fill=green!80!black},
unused/.style={sourcelevel,dashed,opacity=.25}
}

\newcommand*\circled[1]{\tikz[baseline=(char.base)]{
            \node[circled,font=\scriptsize] (char) {#1};}}

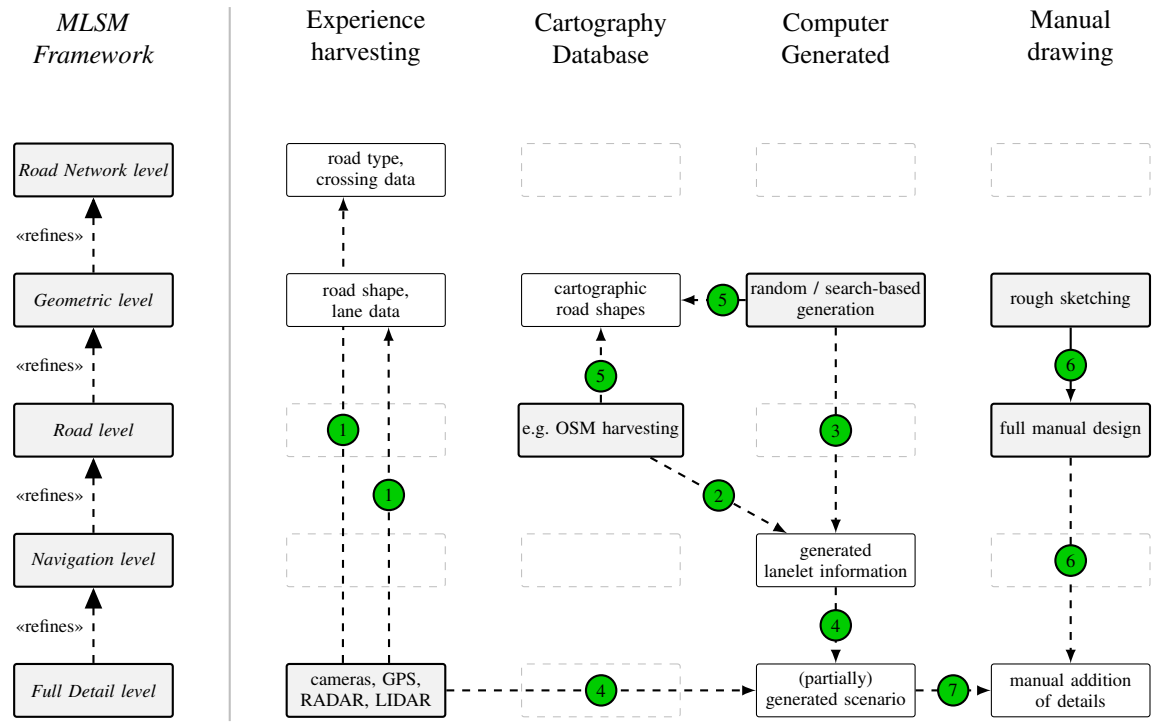
\begin{figure*}[h]
\centering
\begin{tikzpicture}[every node/.style={font=\scriptsize}]

\node[source] (sensors) {Experience\\harvesting};
\node[source,right=1 of sensors] (osm) {Cartography\\Database};
\node[source,right=1 of osm] (generate) {Computer\\Generated};
\node[source,right=1 of generate] (draw) {Manual\\drawing};

\node[source,left=1.5 of sensors] (theoretical) {\emph{\ac{mlsm}}\\\emph{Framework}};
\node[base,below=1 of theoretical] (graph) {\graphlevel};
\node[base,below=1 of graph] (geog) {\geographiclevel};
\node[base,below=1 of geog] (road) {\roadlevel};
\node[base,below=1 of road] (lanelet) {\laneletlevel};
\node[base,below=1 of lanelet] (full) {\fulldetail};

\draw[refines] (geog) -- (graph)   node[pos=.5,left] {<<refines>>};
\draw[refines] (road) -- (geog)    node[pos=.5,left] {<<refines>>};
\draw[refines] (lanelet) -- (road) node[pos=.5,left] {<<refines>>};
\draw[refines] (full) -- (lanelet) node[pos=.5,left] {<<refines>>};

\node[unused] at (sensors |- road) {};
\node[unused] at (sensors |- lanelet) {};

\node[unused] at (osm |- graph) {};
\node[unused] at (osm |- lanelet) {};
\node[unused] at (osm |- full) {};

\node[unused] at (generate |- graph) {};
\node[unused] at (generate |- road) {};

\node[unused] at (draw |- graph) {};
\node[unused] at (draw |- lanelet) {};

\draw[thick,draw=gray,opacity=.5] ($(theoretical.north east)!0.5!(sensors.north west)$) -- ++(0,-9.5);


\node[base] (sensors-full) at (full -| sensors) {cameras, GPS,\\RADAR, LIDAR};

\node[sourcelevel] (sensors-graph) at (sensors |- graph)  {road type,\\crossing data};
\draw[myarrow] (sensors-full.130) -- (sensors-graph.230) node[pos=0.5,circled,solid] {1};

\node[sourcelevel] (sensors-geog) at (sensors |- geog)  {road shape,\\lane data};
\draw[myarrow] (sensors-full.50) -- (sensors-geog.310) node[pos=0.5,circled,solid] {1};

\node[base] (osm-road) at (road -| osm) {\eg \ac{osm} harvesting};
\node[sourcelevel] (generate-lanelet) at  (generate |- lanelet)  {generated\\lanelet information};
\draw[myarrow] (osm-road) -- (generate-lanelet) node[pos=0.5,circled,solid] {2};

\node[base] (generate-geog) at  (generate |- geog)  {random / search-based\\generation};
\draw[myarrow] (generate-geog) -- (generate-lanelet) node[pos=0.5,circled,solid] {3};

\node[sourcelevel] (generate-full) at  (generate |- full)  {(partially)\\generated scenario};
\draw[myarrow] (generate-lanelet) -- (generate-full) node[pos=0.5,circled,solid] (four) {4};
\draw[myarrow] (sensors-full) -- (generate-full) node[pos=0.5,circled,solid] (four) {4};

\node[sourcelevel] (osm-geog) at (osm |- geog) {cartographic\\road shapes};
\draw[myarrow] (generate-geog) -- (osm-geog) node[pos=0.35,circled,solid] {5};
\draw[myarrow] (osm-road) -- (osm-geog) node[pos=0.35,circled,solid] {5};

\node[sourcelevel] (manual-full) at (full -| draw) {manual addition\\of details};
\draw[myarrow] (generate-full) -- (manual-full) node[pos=0.5,circled,solid] {7};

\node[base] (manual-geog) at (geog -| draw) {rough sketching};
\node[base] (manual-road) at (road -| draw) {full manual design};

\draw[myarrow,solid,draw=black] (manual-geog) -- (manual-road) node[pos=0.5,circled,solid] {6} ;
\draw[myarrow] (manual-road) -- (manual-full) node[pos=0.5,circled,solid] {6} ;

\end{tikzpicture}
\caption{Relationships between Scenario sources}
\label{fig:relations}
\end{figure*}

In this section we describe our idea for a \ac{mlsm}, that organises captured scenario information into abstraction levels.
Note that due to space constraints, this paper is limited to the scenarios' map and road structure.
Evidently, the other dimensions (traffic interactions, static objects, weather conditions, occlusions) should also be modelled in a similar, consistent and hierarchical fashion.
For the same reason, our attention is placed on vehicle road, meaning we may omit description of side walks and similar.

\subsection{Abstraction Levels}
In this paper, we consider five hierarchical abstraction levels, building one upon another, increasingly adding further details and concretisation (see \cref{fig:relations}---left).

\paragraph{\graphlevel}
At the top level, a map's road system is viewed as an abstract graph, using road crossings as nodes and roads connecting them (see \cref{fig:graph}).
This viewpoint enables us to pose questions about the style of intersections we might see (3-way / 4-way intersections).
Additionally, important annotations such as the type of road (local road, highway), existence of bike lanes, zebra crossings are added.

\paragraph{\geographiclevel}
The next level adds topological information to the graph, such that we can trace the shape of the road network (see \cref{fig:geographic}). 
This is the level where information about the number of lanes is added.
The level allows us, for instance, to query the general complexity of navigation and evaluate if any lane-changing manoeuvres are potentially required at crossings.
We may also introduce additional nodes, \eg for lane-merging points and similar, if necessary.

\paragraph{\roadlevel}
The \roadlevel signifies the switch from a rather abstract to a more concrete view of the road. 
Here the road shape (\cref{fig:road}) is extended to include specific lane geometries and annotated with road boundary information, pedestrian crossings, etc.
In theory, this level could serve as the operation domain for abstract mission, manoeuvre or motion planners that do not require full information.

\paragraph{\laneletlevel}
On the \laneletlevel the model is extended by particular routing and navigation information. For instance, the Lanelet2~\cite{poggenhans2018lanelet2} library could be used to express where and how a vehicle is allowed to move and switch lanes (see \cref{fig:lanelets}). In the figure, each letter denotes a separate lanelet, which stores its own geometry, reference line, road signs, traffic lights, road markings, etc.
Most modern \ac{ads} simulators should be able to operate at this level of granularity under the assumption that no optical perception (\eg image recognition) is required.

\paragraph{\fulldetail}
On the \fulldetail the model is extended to include all data for reproduction of a scenario map, including very specific information such as road surface type, road damages or potholes, and similar particularities that are abstracted over on other higher levels. 
Casually speaking, this is the level at which a video-game-like rendering is possible.

\subsection{Scenario Modelling Use Cases \& \ac{mlsm}}

The existence of and need for different abstraction levels becomes more evident when we relate them to the individual scenario creation methods and modelling activities.
\cref{fig:relations} splits these activities by the scenario's source. 
Note that the shaded boxes signify the base level at which level the information is initially provided in the source.

For instance, when we use \emph{experience harvesting} to extract scenario information directly ``from the road'', we can often distill very detailed information from cameras, GPS, LIDAR and RADAR sensors, etc.
This means the data can be readily used for capture \& replay simulations or transformed into a scenario using an \ac{sdl}.
Thus, the data is provided at the \fulldetail.
The problem is that despite all the available information, it is rather complex to synthesize general map information from sensors.
Furthermore, any data obtained through reasoning / image analysis is usually only temporarily buffered and erased immediately after use.
Hence, when stored in a database or file store, it is typically impossible to ask questions about road type, road shape or query lane information. 
We believe that this data should be accessible (see \circled{1} in \cref{fig:relations}) for filtering and analysis.

Similarly, when harvesting a map from online cartography services (\eg \ac{gis} databases or \ac{osm}), information is often provided at the road level.
Nonetheless, most \ac{ads} require lanelet or other road routing information for the creation of a valid mission and motion plan.
In the absence of such data, it has to be either manually added, or automatically generated (see \circled{2}), using the cartographic data as a source.

Alternatively, we might use search-based approaches to first create a desired road shape, which is then rendered into a lanelet network according to predefined rules (see \circled{3}).
This approach, situated at the \geographiclevel, was part of one of the 2021 SBST-Competitions\footnote{\url{https://sbst21.github.io/tools/}}, whose goal was to find roads that lead a simulated automated vehicle to drive off the road before reaching the target area.
Competitors used \ac{sbt} methods to generate road shapes that were then enhanced and fed to an \ac{ads} simulator.

Use case \circled{4} shows enhancing a scenario at \laneletlevel to inject realistic sensor information and details such as visual information, road surface and similar for completely realistic simulation and rendering.

Another modelling task we envision is the consistency checking of generated roads.
When dealing with generated information, we typically require confirmation that our data is realistic. 
Thus, for instance, we might compare the generated road shapes to real roads using cartographic information \circled{5}.
Given the complexity of this geometric task, it is vital to operate on an as high abstraction level as possible.

Use case \circled{6} displays the complete manual design of a road, starting at the \geographiclevel or \roadlevel, all the way down to the \fulldetail.
The problem is that this process is very time-consuming, and we envision that typically some parts might be generated using tools such as RoadRunner.
Generally, it might be easier to rely on some form of automated generation techniques, before manually editing to ``smoothen the edges'' and making sure the result matches the particular engineer's expectations and constraints (see \circled{7}).


\section{Scenario Modelling Activities}
\label{sec:activities}

The previous section introduced several use cases and placed them in the schematics of the hierarchical \ac{mlsm} according to their data source.
From our viewpoint, the scenario sources and hierarchical abstraction levels implicitly create a form of grid, such that it should be possible to navigate vertically to the most appropriate abstraction for a given task.
Scenarios should be consistent with one another on the same hierarchical level, independent of the data source.

This section aims to reorganise the above use cases and split them into specific scenario modelling activities. 
Note, that this summary is rather ``academic'' and we expect that, as shown above, for practical applications several activities will be merged and interwoven.

\paragraph{Scenario Querying}
Evidently, \ac{mlsm} has a significant benefit by allowing refined search queries to the scenario model.
Indeed, enabling the filtering of scenarios in a database at the most appropriate level, will lead to more precise search results and enable structured cross-level
queries that combine \eg for number of roads leading to a crossing (\graphlevel), general road shape and lane count (\geographiclevel), roads boundaries and markings, lane widths, traffic signs and regulations (\roadlevel) and specific navigation information (\laneletlevel).

\paragraph{Scenario Adaptation \& Design}
The above described search functionality not only avoids unnecessary complicated searches in scenario databases, 
but also helps scenario designers to efficiently adapt existing scenarios. 
For instance, one might search a database or online cartography network for maps that are similar to the one we might design,
instead of starting from scratch and creating a new one.

\paragraph{Scenario Refinement \& Generation}
By having a clearly defined hierarchy that defines the scenario granularity at each level, 
scenario designers can be guided through the process, making sure that the individual steps are followed.
Furthermore, generation tools can be developed to tackle a specific abstraction level and provide and extend a scenario by the specific information.
For instance, the transition from the \roadlevel to the \laneletlevel requires the analysis of routing information for vehicles and pedestrians.
On the other hand, passing from \graphlevel to \geographiclevel, we might require information such as maximum road curvature, etc.
At each level, it is then possible to perform consistency checks to make sure that the created map asserts correct properties.
When an inconsistency is detected, it is possible to ``step back/up'' a level and adapt at a higher abstraction level.

\paragraph{Requirements Engineering}

Another activity that could benefit from \ac{mlsm} is \emph{requirements engineering} (RE). When discussing about \ac{ads} requirements, a key concept is that of \emph{Operation Driving Domain} (ODD)~\cite{adsWaterlooODDtaxonomy}. 
An ODD specifies the operating conditions in which the \ac{ads} should operate in terms of road characteristics, traffic characteristics, environmental elements, etc. 
Different ODDs can focus on different aspects and may benefit from different representations. 
For example, the \emph{Operational Road Environment Model} (OREM)~\cite{adsWaterlooODDtaxonomy} represents the type of road (\eg number of lanes) in which the ADS should drive, but does not consider elements such as traffic and environmental conditions; as such, a scenario at an intermediate level of abstraction suffices.

In the RE community, the need for managing requirements at different levels of abstraction is widely recognised~\cite{Buhne2004}. Our proposed \ac{mlsm} can be a suitable tool to introduce support for RE activities in the \ac{ads} domain, including specifying use cases for the different types of requirements, using the abstraction level that best matches the type of requirement.

\section{Case study: Search-based testing}\label{sec:caseStudySBT}

The current study has been developed starting from our experience in simulation-based testing of \ac{ads}, and from the issues that we faced in that domain. Therefore, we here provide a more extensive example of application of \ac{mlsm} in this domain. Note that we consider \ac{mlsm} for all scenario dimensions here, as explained in previous sections. 
We therefore assume a hierarchical model for describing \eg other traffic participants' behaviour, the environment, etc.

In simulation-based testing of \ac{ads}, it is known that, in order to thoroughly test some components of the \ac{ads}, reusing existing scenarios (\eg harvested from \ac{osm}) is not enough, and that, instead, scenarios should be \emph{generated}~\cite{hauer2020re}. \Acf{sbt} is a popular technique in which the scenario generation is cast as an optimisation problem, and meta-heuristic search is used to drive the optimisation. The goal is to find scenarios exhibiting different \ac{ads} behaviours, \eg the \ac{av} driving off the lane~\cite{frenetic2021sbst}, the perception component failing to recognise a pedestrian~\cite{Abdessalem2016ASE}, or the \ac{av} colliding with another vehicle~\cite{calo2020Generating}.

\ac{sbt} for \ac{ads} will greatly benefit from the proposed \ac{mlsm}. Indeed, in a classical \ac{sbt} approach, a developer specifies a given search space where to search for scenarios optimising the test goal: this usually consists in selecting a particular map, identifying some traffic participants (other vehicles and pedestrian), and parametrising some elements of these participants (\eg initial position and speed) that define the search space. It is apparent that the effectiveness of \ac{sbt} greatly depends on the definition of the search space, that can only be defined at the most detailed scenario level.

We envision that \ac{mlsm} enables the search to explore more diverse scenarios and thus to cover more \ac{ads} behaviours. Indeed, the search space could be defined at the different levels of abstraction provided by \ac{mlsm}; the search algorithm could take advantage of this representation and navigate, in a Monte Carlo Tree Search fashion, through the different abstraction levels. Let us consider, as an example, the search task of finding a scenario in which the \ac{av} drives off the road. The search algorithm could start evaluating different road maps at the higher level of abstraction, to estimate in which ones the \ac{av} has the possibility of driving off the lane, \eg abstract scenarios with a given degree of curvature. Starting from a promising scenario $\mathit{S}$ selected at the most abstract level, the search could continue exploring refined scenarios derived from $\mathit{S}$, containing other traffic participants, traffic signals, and so on. In these refined scenarios, the search could find the one in which the \ac{av} indeed drives off the road: for example, the search could find a scenario in which the appearance of a preceding vehicle during a turn leads the \ac{av} to suddenly break and lose control.

Note that the number of refined search spaces could be even higher. Let us assume that in the scenario found in the last step, the \ac{av} does not actually go out of the road, but just gets close to the boundary. The search space could be further refined by adding environmental elements to the search, such as rain or fog. This means, the search would also operate on another scenario dimension's \ac{mlsm}, finding that adding some fog to the previous scenario can delay the moment in which the \ac{av} recognises the preceding vehicle, so leading to an even sharper evasion manoeuvre that, in this case, leads it off the road.

\section{Future Work: Towards an Implementation}
\label{sec:implementation}

At the moment, our work on the \ac{mlsm} framework is of theoretical nature. 
In collaboration with our colleagues and collaborators from within the \ac{ads} domain, we are evaluating the requirements and trying to find inconsistencies and ambiguities.
Our next steps, include the finalisation of a more refined version of the framework for all five scenario dimensions.
Furthermore, we are evaluating possible implementation technologies that are compatible with currently existing standards, \acp{sdl} and scenario description customs.

Our project includes the need to overcome a great deal of diversity in the field, including the different syntax and semantics of \acp{sdl},
different scenario description paradigms (descriptive, prescriptive), and the fact that the \ac{ads} domain is fast-paced with continuous release of new and change of existing standards and formats.

Currently, we try to explore, exploit and reuse existing techniques by \eg building a scenario's map description on well-known \ac{gis}-standards. 
This system already allows the definition of layered information, data import/export and the definition of constraints.
Nonetheless, advanced use cases such as constraint checking, need to be efficiently integrated.
We hope that our good relations to the modelling community will help us overcome (some of) the complexity and use the experience present in the domain.
This includes the evaluation of \eg domain-specific constraint languages, languages and meta-models for description of vehicle behaviour, and experiences in model serialisation and designing model databases.

\glsresetall

\section{Conclusion}
This paper raises an important issue present in current \ac{ads} scenario modelling, which typically leads to scenarios only being defined at the most fine-grained level of abstraction, using as much detail as possible.
In the process of capturing this data, high-level information at higher abstraction levels is either not captured in the first place or removed.
While a large amount of detailed information is important for realistic simulation, some tasks such as searching of scenarios in databases (\eg by road shape) or consistency checking of generated scenarios are rendered unnecessarily complex or even impossible, as the data is at a too low abstraction level.
Our work describes the need for a hierarchical \ac{mlsm} that enables modelling at the most appropriate level of abstraction, and describes how \ac{mlsm} can benefit the \ac{ads} community.


\bibliographystyle{IEEEtran}
\bibliography{bibliography}

\end{document}